# Dynamics of fixed-volume pinned films – dealing with a non-self-adjoint thin film problem

Israel Gabay[1], Vesna Bacheva[2,1], Dotan Ilssar[3], Moran Bercovici[1], Antonio Ramos[4], Amir Gat[1]

[1] *Faculty of Mechanical Engineering, Technion – Israel Institute of Technology, Haifa, Israel*
[2] *IBM Research Europe, Zurich, Switzerland*
[3] *Department of Mechanical and Process Engineering, ETH Zürich, Zürich 8092, Switzerland*
[4] *Depto. Electronica y Electromagnetismo, Facultad de Fisica, Universidad de Sevilla, Sevilla, Spain*

**Abstract**

The use of thin liquid films has expanded beyond lubrication and coatings, and into applications in actuators and adaptive optical elements. In contrast to their predecessors, whose dynamics can be typically captured by modelling infinite or periodic films, these applications are characterized by a finite amount of liquid in an impermeable domain. The global mass conservation constraint, together with common boundary conditions (e.g., pinning) create quantitatively and qualitatively different dynamics than those of infinite films. Mathematically, this manifests itself as a non-self-adjoint problem. This work presents a combined theoretical and experimental study for this problem. We provide a time-dependent closed-form analytical solution for the linearized non-self-adjoint system that arises from these boundary conditions. We highlight that, in contrast to self-adjoint problems, here special care should be given to deriving the adjoint problem to reconstruct the solution based on the eigenfunctions properly.

We compare these solutions with those obtained for permeable and periodic boundary conditions, representing common models for self-adjoint thin-film problems. We show that while the initial dynamics are nearly identical, the boundary conditions eventually affect the film deformation as well as its response time. To experimentally illustrate the dynamics and to validate the theoretical model, we fabricated an experimental setup that subjects a thin liquid film to a prescribed normal force distribution through dielectrophoresis, and used high-frame-rate digital holography to measure the film deformation in real-time. The experiments agree well with the model and confirm that confined films exhibit different behaviour which could not be predicted by existing models.

## 1. Introduction

The deformation of thin liquid films has been investigated for more than a century (Rayleigh, 1916; Bénard, 1900), owing to their importance in a wide range of natural phenomena (Fink et al., 1990; Hewitt et al., 2015; Simpson, 1982; Grotberg et al., 2004; McGraw et al., 1996) and engineering applications (Oron et al., 1997; Zheng, Fontelos, Shin, and Stone, 2018; Backholm et al., 2014; Craster et al., 2009). To date, solutions to the thin film problem have focused on boundary conditions describing infinite films (Deissler et al., 1992; Gjevik, 1970), periodic films (Chappell et al., 2020; Frumkin et al., 2016; Williams et al., 1982), or those that give rise to self-similar solutions (Zheng, Fontelos, Shin, and Stone, 2018; Backholm et al., 2014; Zheng, Fontelos, Shin, Dallaston, et al., 2018; Dallaston et al., 2017). Surprisingly, despite the fact that a large number of applications and phenomena involve films that are pinned on impermeable boundaries, analytical modelling of this fundamental case has been overlooked. While existing solutions provide general insight into thin films dynamics, the non-penetration conditions at the boundaries of finite domains, together with pinning conditions, eliminate the self-adjointness of the governing equations and alter the dynamic response of the system in a manner that existing models fail to capture.

Among the engineering applications of thin films in close domains, one of the most active fields of research is adaptive optics where interfacial deformations can be leveraged to create smooth optical elements (Brown et al., 2009; Banerjee et al., 2018; Eshel et al., 2021; zhao et al., 2021). Another emerging application is in the field of reconfigurable microfluidics (Paratore et al., 2022), where thin liquid films could be used to dynamically deform the surface of a thin membrane of a microfluidic chip, in order to control its functionality in real time. The film dynamics in such applications can be described by the same set of equations, whether a liquid-fluid or a liquid-membrane interface is used (Hosoi et al., 2004; Kodio et al., 2017; Boyko et al., 2019; 2020). The ability to design such devices, predict their performance, and understand their fundamental limitations, would greatly benefit from a theoretical framework that allows their analysis.

We here present a combined theoretical and experimental study of the dynamics of a thin liquid film that is pinned at the boundaries of an impermeable finite domain, and subjected to a normal stress distribution at its liquid-air interface. We provide a time-dependent closed-form analytical solution for the linearized evolution problem, which is non-self-adjoint. We compare this solution with existing solutions for the same configuration but with self-adjoint boundary conditions and show that while the initial dynamics are identical, the difference in boundary conditions quickly affects the entire domain, and the response time of the system. To experimentally illustrate the dynamics and to validate the model, we developed a setup that enables high frame-rate measurements of microscale deformations based on digital holography. We use this setup to investigate the deformation of a thin liquid film actuated by an array of electrodes imposing a dielectrophoretic (DEP) force on the interface (Gabay et al., 2021), showing good agreement to the theory.

## 2. Analytical model

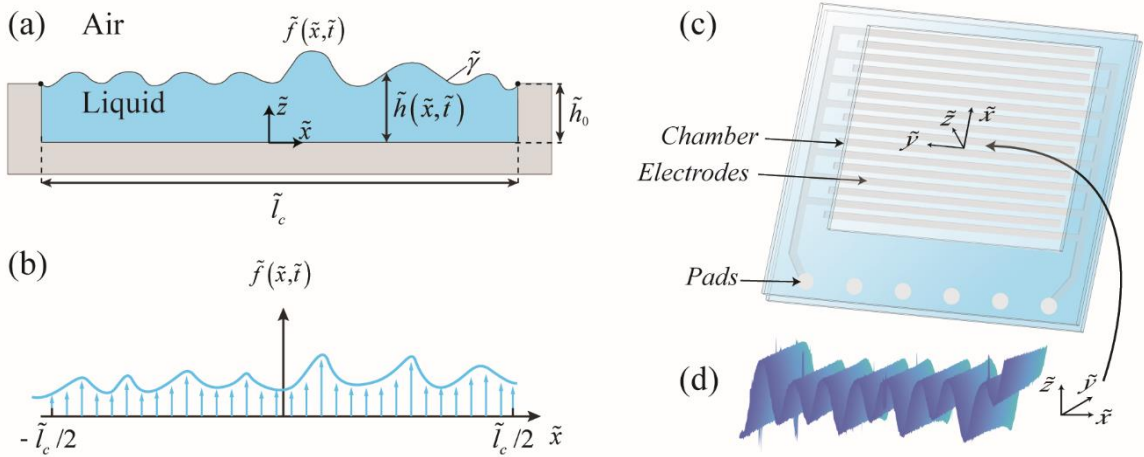

*Figure 1. (a) Two-dimensional illustration of the investigated model comprised of a shallow chamber of length $\tilde{l}_c$ and height $\tilde{h}_0$ filled with a liquid of mass density $\tilde{\rho}$, dynamic viscosity $\tilde{\mu}$, and open to the air above. The liquid-air interface with surface tension $\tilde{\gamma}$ is subjected to a normal force distribution $\tilde{f}(\tilde{x},\tilde{t})$, resulting in a deformed liquid height $\tilde{h}(\tilde{x},\tilde{t})$. At the edges of the chamber the liquid is pinned and cannot penetrate the solid, resulting in global mass conservation. (c) An illustration of a microfabricated chamber used in the experiments. The bottom of the chamber is patterned with an array of electrodes imposing a dielectrophoretic (DEP) force on the liquid-air interface. (d) Example of an experimentally observed deformation of the liquid film when subjected to a DEP force distribution.*

Figure 1 presents an illustration of the investigated configuration. Consider a shallow chamber of depth $\tilde{h}_0$ and length $\tilde{l}_c$ filled with a Newtonian liquid of mass density $\tilde{\rho}$, dynamic viscosity $\tilde{\mu}$ and volume $\tilde{V}$. The liquid-air interface has a surface tension $\tilde{\gamma}$ and is subjected to a normal force distribution $\tilde{f}(\tilde{x},\tilde{t})$ that induces a spatio-temporal deformation of the interface, $\tilde{h}(\tilde{x},\tilde{t})$. Under the assumptions of shallow geometry $(\tilde{h}_0 \ll \tilde{l}_c)$ and negligible fluid inertia, the Navier-Stokes equations reduce to the lubrication equations,

(2.1)
$$\tilde{u}_{\tilde{x}} + \tilde{w}_{\tilde{z}} = 0, \quad \tilde{p}_{\tilde{x}} = \tilde{\mu} u_{\tilde{z}\tilde{z}}, \quad \tilde{p}_{\tilde{z}} = -\tilde{\rho}\tilde{g},$$

where $(\tilde{u}, \tilde{w})$ are the velocity component in the $\tilde{x}$ and $\tilde{z}$ directions respectively, and $\tilde{p}$ is the fluidic pressure. We marked the derivative with respect to $x$ of a multivariable function, $g(x,z)$, as $g_x$. Requiring no-slip at the bottom surface, a stress-balance at the free surface, and utilizing the kinematic boundary condition, the equations can be further simplified to the Reynolds equation, (Oron et al., 1997; Leal, 2007)

(2.2)
$$\tilde{h}_{\tilde{t}} = \frac{1}{3\tilde{\mu}} \left( \tilde{h}^3 \tilde{p}_{\tilde{x}} \right)_x.$$

Under the longwave approximation (linear curvature), the fluidic pressure, $\tilde{p}$, is given by

(2.3)
$$\tilde{p} = \tilde{\rho}\tilde{g}(\tilde{h} - \tilde{z}) - \tilde{\gamma}\tilde{h}_{\tilde{x}\tilde{x}} - \tilde{f} + \tilde{C},$$

where the first term on the right-hand-side is the hydrostatic pressure, the second one is the capillary pressure associated with the curvature of the interface, the third one is the normal stress distribution on the interface, and $\tilde{C}$ is a constant. We denote the film deformation as $\tilde{d}(\tilde{x},\tilde{t}) = \tilde{h}(\tilde{x},\tilde{t}) - \tilde{h}_0$, and define the non-dimensional quantities $t = \tilde{t}/\tilde{\tau}$, $x = \tilde{x}/(\tilde{l}_c/2)$, $d = \tilde{d}/\tilde{h}_0$, $f = \tilde{f}/\tilde{F}$. By further assuming small deformations, $d \ll 1$, we obtain the non-dimensional linearized evolution equation for the deformation,

(2.4)
$$d_t + d_{xxxx} - \text{Bo} \cdot d_{xx} = -\Pi f_{xx},$$

where $\text{Bo} = \frac{\tilde{\rho}\tilde{g}\tilde{l}_c^2}{4\tilde{\gamma}}$ is the Bond number, $\Pi = \frac{\tilde{F}\tilde{l}_c^2}{4\tilde{\gamma}\tilde{h}_0}$ represents the non-dimensional interfacial force magnitude, and the resulting time scale is $\tilde{\tau} = \frac{3\tilde{\mu}\tilde{l}_c^4}{16\tilde{h}_0^3\tilde{\gamma}}$.

At the boundaries we require the liquid to be pinned,

(2.5)
$$d(x = \pm 1, t) = 0,$$

as well as no penetration through the boundaries, i.e., $u(x = \pm 1, t) = 0$. Integrating the x-momentum equation (2.1) with respect to $\tilde{z}$, yields $\tilde{u} = \tilde{p}_{\tilde{x}} \frac{1}{2\tilde{\mu}} \left( \frac{\tilde{z}^2}{\tilde{h}^2} - 2\frac{\tilde{z}}{\tilde{h}} \right)$, which translates the no penetration boundary conditions to $\tilde{p}_{\tilde{x}} = 0$ at the chamber edges. Applying this requirement to equation (2.3), yields the following boundary condition on the deformation,

(2.6)
$$d_{xxx}(x = \pm 1, t) - \text{Bo} \cdot d_x(x = \pm 1, t) = -\Pi f_x(x = \pm 1, t).$$

These boundary conditions are accompanied by an arbitrary initial condition, $d(x, t = 0) = d_0(x)$.

A straightforward approach for solving the system (2.4) with boundary conditions (2.5) and (2.6) is by separation of variables. However, as we detail here, this set of boundary conditions is responsible for transforming the system into a non-self-adjoint one, requiring special treatment in the separation of variables procedure. Equations (2.7) - (2.11) follow the standard procedure for separation of variables, yet we present it explicitly for completeness and consistency with the derivation in equations (2.12) - (2.16), where we construct the solution based on the adjoint operator.

We first homogenize the boundary conditions by expressing the deformation as $d(x,t) = \hat{d}(x,t) + P(x,t)$, $P(x,t)$ is the homogenising function that should be selected as a polynomial of sufficient degree in $x$ such that the source terms will vanish from all boundary conditions, yielding the system

(2.7)
$$\begin{aligned}
&\hat{d}_t + \hat{d}_{xxxx} - \text{Bo}\,\hat{d}_{xx} = -P_t - \Pi f_{xx} - P_{xxxx} + \text{Bo}\,P_{xx} \\
&\hat{d}(x = \pm 1, t) = 0 \\
&\hat{d}_{xxx}(x = \pm 1, t) - \text{Bo} \cdot \hat{d}_x(x = \pm 1, t) = 0 \\
&\hat{d}(x, t = 0) = d_0(x) - P(x, 0)
\end{aligned}$$

A compact and convenient homogenising function can be constructed from a 3$^{\text{rd}}$ order polynomial, $P(x,t) = \Pi \dfrac{f_x(1,t) - f_x(-1,t)}{4\text{Bo}}(x^2 - 1) + \Pi \dfrac{f_x(1,t) + f_x(-1,t)}{4(\text{Bo} - 3)}(x^3 - x)$, which is valid for any Bond number except $\text{Bo} = 0, 3$. To cover these two specific cases, a 4$^{\text{th}}$ order polynomial can be constructed.

Next, by separation of variables, we suggest a solution in the form of $\hat{d}(x,t) = X(x)T(t)$ and substitute it into the homogeneous form of equation (2.7), which yields,

(2.8)
$$-\frac{T'}{T} = \frac{X^{(4)}}{X} - \text{Bo}\,\frac{X''}{X} = \lambda^4.$$

The eigenvalue problem, (2.8), followed by the set of boundary conditions,

(2.9)
$$X(\pm 1) = 0, \quad X'''(\pm 1) - \text{Bo}\,X'(\pm 1) = 0,$$

where each prime denotes a derivation with respect to the single variable of the function. The solution of the system (2.8) and (2.9) is given by

(2.10)
$$X_n = \begin{cases} \dfrac{\sqrt{\text{Bo}}\left(\cosh\left(\sqrt{\text{Bo}}\,x\right) - \cosh\left(\sqrt{\text{Bo}}\right)\right)}{2\left(\sinh\left(\sqrt{\text{Bo}}\right) - \sqrt{\text{Bo}}\cosh\left(\sqrt{\text{Bo}}\right)\right)} & \text{for } n = 0 \\ \sin(\alpha_n x) - \dfrac{\sin(\alpha_n)}{\sinh(\beta_n)}\sinh(\beta_n x) & \text{for odd } n \text{ numbers} \\ \cos(\alpha_n x) - \dfrac{\cos(\alpha_n)}{\cosh(\beta_n)}\cosh(\beta_n x) & \text{for even } n \text{ numbers} \end{cases}$$

where $\alpha_n = \sqrt{\dfrac{-\text{Bo} + \sqrt{\text{Bo}^2 + 4\lambda_n^4}}{2}}$, and $\beta_n = \sqrt{\dfrac{\text{Bo} + \sqrt{\text{Bo}^2 + 4\lambda_n^4}}{2}}$. The eigenvalues are calculated from two transcendental equations, one for each set of eigenfunctions respectively,

(2.11)
$$\begin{cases} \alpha \tan(\alpha) = -\beta \tanh(\beta) \\ \beta \tan(\alpha) = \alpha \tanh(\beta) \end{cases}.$$

We note that all eigenfunctions have a zero spatial integral, except for the zero eigenfunction that corresponds to a steady state solution for a non-actuated system, and its integral sets the volume in the chamber.

The eigenfunctions (2.10) are non-orthogonal to one another, which is an indication of the non-self-adjointness of the system. Nonetheless, we can proceed by using the biorthogonality relation of the system with its adjoint system, $\langle L[X], Y \rangle = \langle X, L^+[Y] \rangle$, where $\langle U, V \rangle = \int_{-1}^{1} UV dx$ is the basic inner product, $L$ is the original differential operator, $L^+$ is the Lagrange adjoint differential operator, and $X$ and $Y$ are their respective solutions. We define the spatial differential operator from equation (2.8) as $L[X] = X^{(4)} - \text{Bo} X''$. Since this spatial differential operator has constant real coefficients and only even derivatives, we are guaranteed that the Lagrange adjoint operator, $L^+$, is identical to $L$ (Coddington et al., 1955). Explicitly expanding the biorthogonality relation while integrating by parts and using the boundary conditions (2.9), yields

(2.12) $$[X''Y' + X'Y'']_{-1}^{1} = 0,$$

which can hold only if $Y'(\pm 1) = Y''(\pm 1) = 0$, thus defining the complementary boundary conditions for $Y$. These conditions are clearly different from those for $X$, (2.9), and thus the problem is not self-adjoint.

The solution for the adjoint system is given by,

(2.13) $$Y_n = \begin{cases} 1 & \text{for } n = 0 \\ \sin(\alpha_n x) - \dfrac{\alpha_n \cos(\alpha_n)}{\beta_n \cosh(\beta_n)} \sinh(\beta_n x) & \text{for odd } n \text{ numbers} \\ \cos(\alpha_n x) + \dfrac{\alpha_n \sin(\alpha_n)}{\beta_n \sinh(\beta_n)} \cosh(\beta_n x) & \text{for even } n \text{ numbers} \end{cases}.$$

The transcendental equations for the two sets of eigenvalues (for odd and even $n$) are identical to those of the original system, and thus the eigenvalues are the same. The solution to the deformation can now be constructed by substituting the separation of variables expression, i.e., $\hat{d}(x,t) = \sum_{n=0}^{\infty} X_n(x) T_n(t)$, to the non-homogenous equation, (2.7), yielding

(2.14) $$\sum_{n=0}^{\infty} X_n(x) T_n'(t) + \sum_{n=0}^{\infty} X_n^{(4)}(x) T_n(t) - \text{Bo} \sum_{n=0}^{\infty} X_n''(x) T_n(t) = \sum_{n=0}^{\infty} X_n(x) a_n(t) + \sum_{n=0}^{\infty} X_n(x) b_n'(t),$$

where $-\Pi f_{xx} + \text{Bo} \cdot P_{xx} - P_{xxxx} = \sum_{n=0}^{\infty} a_n(t) X_n(x)$, $P(x,t) = \sum_{n=0}^{\infty} b_n(t) X_n(x)$, $d_0 - P(x,0) = \sum_{n=0}^{\infty} I_n X_n(x)$.

Multiplying equation (2.14) by $Y_m$ and integrating over the domain with respect to $x$, as well as using the initial condition in (2.7), yields

(2.15) $$T_n = \int_{s=0}^{t} e^{\lambda_n^4 (s-t)} \left[ a_n(s) - b_n'(s) \right] ds + e^{-\lambda_n^4 t} I_n.$$

Thus, the solution for the deformation is given by $d(x,t) = \sum_{n=0}^{\infty} T_n(t) X_n(x) + P(x,t)$.

For a general function $g(x)$ the series expansion by the system eigenfunctions is given by,

$$(2.16) \quad g(x)=\sum_{n=0}^{\infty} c_n X_n(x); \quad c_n = \frac{\int_{-1}^{1} g(x) Y_n dx}{\int_{-1}^{1} X_n Y_n dx}.$$

We note that since our system has real and constant coefficients as well as homogenous boundary conditions, and since its eigenvalues are real with no duplicity, we are guaranteed that the eigenfunctions of both the original and the adjoint systems form a complete set (Coddington et al., 1955).

## 3. Dynamics of the non-self-adjoint closed-chamber system and comparison with self-adjoint periodic and open-chamber systems

In this section we present the behaviour of a thin liquid film within a closed chamber and compare it to classical solutions for self-adjoint configurations such as periodic and open-chamber problems (solutions for the self-adjoint problems in appendix A). Table 1 lists the boundary conditions for each of these cases. In the closed chamber problem ('Closed') the liquid is pinned at both edges and the side walls of the chamber are impermeable. In the open chamber problem ('Open') the liquid is also pinned, but is free to flow in and out of the chamber (i.e., the total volume in the chamber can vary in time). In the periodic case ('Periodic'), there is no physical chamber, but the boundary conditions define a domain where the film deformation and its derivatives at one end are equal to their values at the other end. For simplicity, we consider here the case of Bo=0, but solutions for other Bond values are readily available and do not introduce a fundamental change in behaviour.

*Table 1 – Summary of the three systems considered in this section.*

| Terminology | Illustration | Boundary conditions | | |
|---|---|---|---|---|
| Closed | 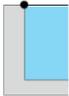 | pinning impermeable | 1,2) $d(\pm 1,t)=0$ <br> 3,4) $d_{xxx}(\pm 1,t) - \text{Bo} \cdot d_x(\pm 1,t) = -\Pi f_x(\pm 1,t)$ | | |
| Open | 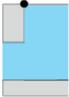 | pinning permeable | 1,2) $d(\pm 1,t)=0$ <br> 3,4) $d_{xx}(\pm 1,t) = -\Pi f(\pm 1,t)$ | | |
| Periodic | 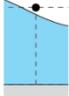 | periodicity | 1) $d(-1,t)=d(1,t)$    2) $d_x(-1,t)=d_x(1,t)$ <br> 3) $d_{xx}(-1,t)=d_{xx}(1,t)$    4) $d_{xxx}(-1,t)=d_{xxx}(1,t)$ | | |

Figure 2 presents the solutions of the deformation for all three systems, at different times, for a harmonic ($f=0.5[1-\cos(10\pi x)]$), and for a gaussian ($f=\exp[-0.5(x+0.85)^2/0.05^2]$) force distribution with an initially flat interface, $d_0(x)=0$. At early times, for both actuations, the deformation of the liquid-air interface is roughly the same for all systems. However, with time, deviations begin to emerge from the boundaries inward.

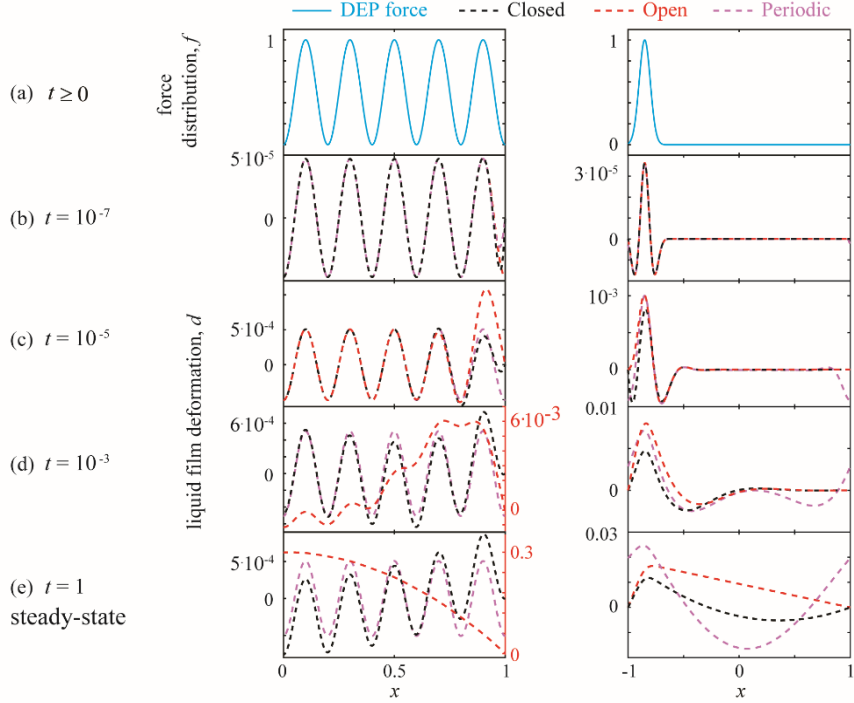

*Figure 2. The effect of different sets of BCs on the film dynamics for harmonic (left column, l) and localised (right column, r) force distributions. (a) The external force distributions applied on the liquid-air interface along the chamber. (b) - (e) Comparison of the analytical solutions for the case of pinned film and impermeable boundaries ('Closed chamber', black dashed line), pinned film and permeable boundaries ('Open chamber', red dashed line), and 'Periodic' (light purple dashed line), at different times. All sub-figures in the left column present half of the chamber, with a symmetry line at x=0. At early times, the deformation exhibited by all systems is nearly identical. With time, deviations start evolving from the boundaries inward. At the harmonic forcing case, since the force is strictly positive, the Open system results in a continuous influx of liquid which ultimately leads to inflation of the interface to a 'balloon'-like structure. In contrast, the Closed and the Periodic systems, owing to their fixed mass, maintain short-wave deformations which follow the force distribution even at steady state, as shown in sub-figure (e.l). The film deformation of the Open system at late times is much larger than the deformation of the other two systems. Thus, sub-figures (d.l) and (e.l) present two different y axis, left for Closed and Periodic systems, and right for the Open system. In the Gaussian forcing case the Open and the Closed systems show similar behaviour due to the pinning condition (although the latter is not a fixed volume system) in contrast to the Periodic system. The solution presented for all systems is for the case of an initially flat interface, $d_0(x) = 0$, and $\text{Bo} = 0$.*

We note that the harmonic force is strictly positive. In the Open case, which allows for liquid influx, we see that this positive force results in 'ballooning' of the film and dominance of the lowest wavenumber. In contrast, in the Closed and Periodic systems, mass conservation does not allow for such inflation and the deformation follows the wavenumber of the force. Another key difference between the systems is that the amplitude of the deformation in the Open system is significantly larger (note the two different y-axes in subfigures 2.d.l, 2.e.l). For the localized force, the deformation in the Closed system resembles that of the Open system (though the former is mass conserving, and the latter is not). The most significant difference is seen here in comparison with the Periodic system in which the film height at the boundary is not fixed, resulting in much greater deformation (albeit mass conserving, i.e., both positive and negative deformations are accentuated). Thus, we can see from the results that neither the Open nor the Periodic formulations can correctly model the dynamics of a Closed system.

Figure 2 shows that the Open, Closed, and Periodic systems reach a steady state at different times. In Figure 2, the localized actuation (right column) triggers a range of system eigenfunctions, and the time to steady state scales as $\lambda^{-4}$, where $\lambda$ is the first (slowest) eigenvalue. The eigenvalues of the system are of course dependent of its boundary conditions. For example, at $Bo=0$, the time scales for steady state in the Open, Closed, and Periodic systems are respectively $\tilde{\tau} \cdot (2/\pi)^4$, $\tilde{\tau} \cdot (1/2.37)^4$, and $\tilde{\tau} \cdot (1/\pi)^4$. i.e. the Closed system is 5-fold faster than the Open system, but 3-fold slower than the Periodic system. Thus, solving for the accurate eigenvalues, meaning using the correct boundary conditions of the system is crucial in order to properly predict the system dynamics.

## 4. Effect of Bond number on system dynamics: response time and deformation magnitude

In this section we investigate the relation between the Bond number and both the response time of the system and the magnitude of the deformation at steady state. The liquid pressure is constant at steady state, thus substituting the non-dimensional parameters defined in (2.4) into equation (2.3) yields

$$d_{xx} - Bo \cdot d + \Pi \cdot f = C, \qquad (4.1)$$

where $C$ is a constant that depends on the liquid volume. For the case of large Bond numbers, $Bo \gg 1$, the deformation can be written as $d \approx \Pi f / Bo + const$. We note that this equation holds only sufficiently far from the chamber edges; close to the edges, the requirement for mass conservation together with pinning conditions may result in significant curvatures, i.e., no negligible $d_{xx}$.

While $\tilde{\tau}$ was conveniently defined in (2.4) as being independent of the Bond number, the response time of the system is in fact strongly influenced by it, as indicated by the relation between the eigenvalues and the Bond number in (2.10). This influence is further enhanced through the quartic dependence of the timescale on the system eigenvalues, $\lambda_n$. The eigenvalues can be obtained by solving a pair of transcendental equations (2.11), for the odd and the even eigenvalues,

$$(4.2.a) \quad \sqrt{\frac{-Bo+\sqrt{Bo^2+4\lambda_n^4}}{Bo+\sqrt{Bo^2+4\lambda_n^4}}} \tan\left(\sqrt{\frac{-Bo+\sqrt{Bo^2+4\lambda_n^4}}{2}}\right) = -\tanh\left(\sqrt{\frac{Bo+\sqrt{Bo^2+4\lambda_n^4}}{2}}\right),$$

$$(4.2.b) \quad \sqrt{\frac{-Bo+\sqrt{Bo^2+4\lambda_n^4}}{Bo+\sqrt{Bo^2+4\lambda_n^4}}} \tanh\left(\sqrt{\frac{Bo+\sqrt{Bo^2+4\lambda_n^4}}{2}}\right) = \tan\left(\sqrt{\frac{-Bo+\sqrt{Bo^2+4\lambda_n^4}}{2}}\right).$$

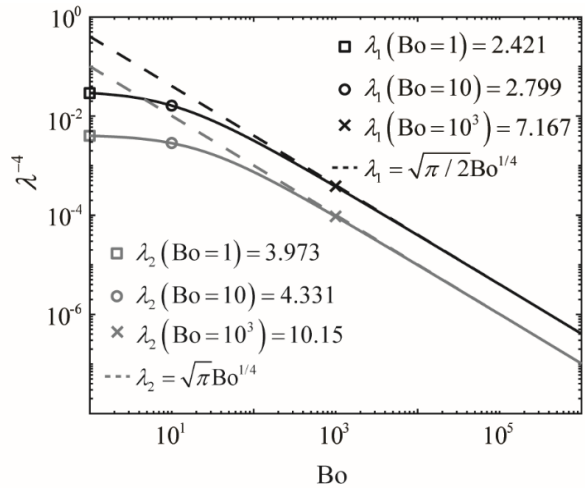

*Figure 3. The eigenvalues of the system as a function of the Bond number. The response time of the system is proportional to, which for large Bo numbers is inversely proportional to* Bo. *Thus, as* Bo *increases, e.g., by increasing the density of the liquid or increasing the size of the container, the response time of the system can shorten significantly.*

The solid lines in Figure 3 present the dependence of the resulting non-dimensional response time, $\lambda^{-4}$, on Bo. For large Bond numbers the asymptotic solutions for the eigenvalues are simply $\lambda_1 = \sqrt{\frac{\pi}{2}} Bo^{1/4}$, $\lambda_2 = \sqrt{\pi} Bo^{1/4}$, and are indicated by the black and the grey dashed lines respectively. For clarity, the square, circle, and x markers indicate the timescale at specific Bond numbers listed in the figure's legend. Clearly, the effect of Bo can be quite significant; the same system at Bo=0 would have a response time 84-fold slower than the same system at $Bo = 1000$.

## 5. Experimental validation

### 5.1. Experimental setup

To experimentally study the dynamics of thin film deformation, we fabricated a shallow fluidic chamber with embedded electrodes which can impose a dielectrophoretic force on the interface of a thin liquid film resting in the chamber. The chamber is 4 mm long, 8 mm wide and 50 μm deep (as shown in subfigure 4a). We filled it with 1.6 μl of silicone oil which has mass density of $\tilde{\rho} = 970$ kg/m$^3$, surface tension of $\tilde{\gamma} = 21 \cdot 10^{-3}$ N/m and electric permittivity of $\tilde{\varepsilon} = 2.7\tilde{\varepsilon}_0$, where $\tilde{\varepsilon}_0$ is the permittivity of free space. The bottom of the chamber is patterned with an array of 200 μm wide parallel electrodes, gapped by 200 μm edge-to-edge. By introducing a sinusoidal electric potential difference to the electrodes (500 Vpp, 10 kHz), the resulting localized electric field imposes dielectrophoretic (DEP) forces on the oil-air interface and deforms it. Figure 4b presents the digital holographic microscope (Cuche et al., 1999) we used to observe the time-dependent deformation of the interface. For each experiment, we initially acquire the baseline surface topography prior to activation of the electric field, and then subtract it from all subsequent measurements.

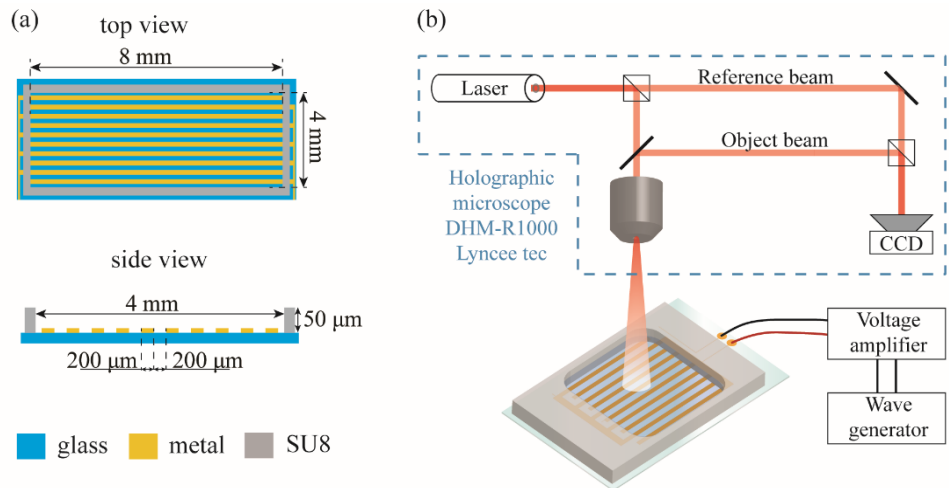

*Figure 4. Illustration of the fluidic chamber the experimental setup. (a) The resulted fluidic chamber configuration and it's fabrication process. At first, the electrodes created by a lift-off process using a 2 nm thick Pt layer sandwiched between two 2 nm thick Ti layers on a borosilicate wafer which forms an array of interdigitated electrode configuration with 200 μn wide electrodes, gapped by 200 μm edge to edge. Second, the chamber's walls created by patterning a 50 μm thick layer of SU-8 photoresist on top of the wafer. Finally, we diced the wafer to separate individual devices. (b) Schematics illustration of the experimental setup. The device is mounted on a home-made connector composed of 3D printed housing and a printed circuit board (PCB) containing electric pins interfacing the device with a voltage amplifier (2210-CE, TREK), which amplifies the voltage output of a wave generator (TG5012A, AIM-TTI Instruments). The device is placed under a digital holographic microscope (DHM R1000, Lyncee tec) allowing observation of the oil-air interface by recording a hologram image on a digital sensor and using a numerical algorithm for real-time reconstruction.*

## 5.2. Experimental results

Figure 5 compares the experimental measurements with an analytical solution based on equation (2.4). The analytical solution is obtain using the force distribution presented in subfigure 5a, corresponding to the DEP force on a liquid-air interface from periodic electrodes (Gabay et al., 2021). Subfigures 4b-f present the evolution of the film in time, showing the gradual appearance of the different spatial modes. At very early times (0.2 ms) the highest spatial frequency is dominant, and the 'tooth'-shaped peaks of the force distribution are also clearly observable in the analytical solution for the deformation. At 2 ms, the model and the experimental measurements agree well and show that the $10^{th}$ even mode of the system has grown significantly (note the change in scale from nm to μm) and dominates the solution. The amplitude of the periodic deformation steadily increases, until at 37 ms a new qualitative change appears, initially in the form of a larger amplitude wave that penetrates from the boundaries. This is predicted well by the analytical model in good agreement with the experiments. By 3.55 s, the entire central region of the liquid has risen at the expense of the edges. The results continue to show good agreement, though some asymmetry begins to appear in the experimental measurements, resulting in slightly larger quantitative deviations. This is likely due to an additional low spatial frequency of the force distribution in the experiment that is not captured by the analytical force model, and which therefore appears only at long times. At 100 s the system has achieved its steady state. The deformation at the centre of the chamber is strictly positive. The overall shape and multitude of spatial frequencies is captured well by the theory, though quantitative differences, particularly on the right side of the chamber, are now evident.

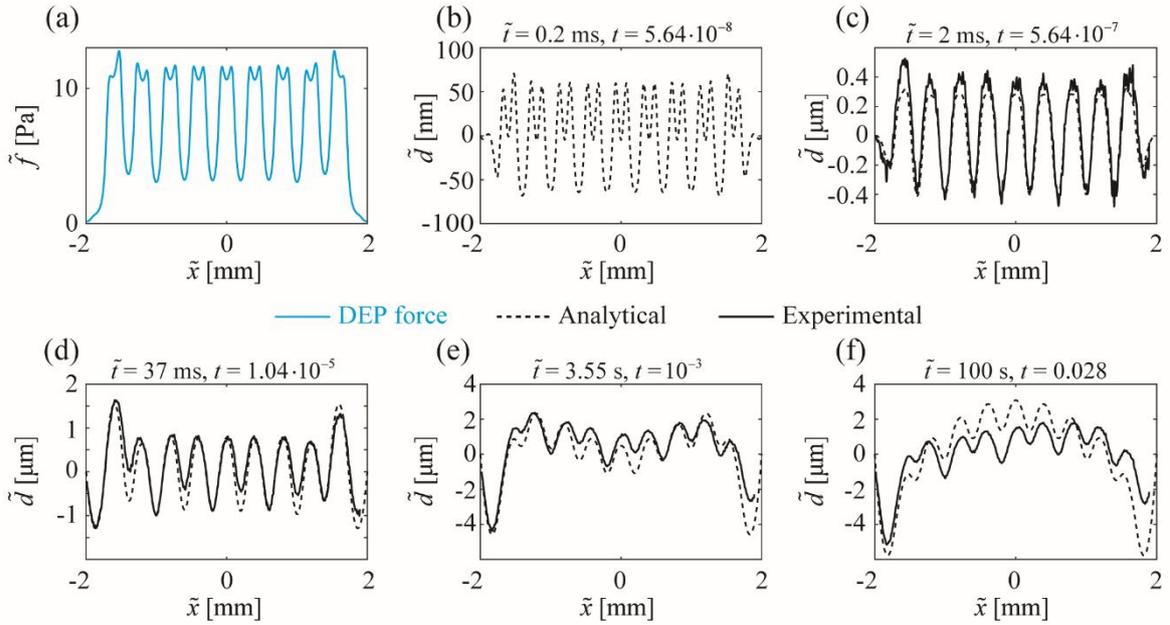

*Figure 5. Validation of the analytical model against experimental measurements. (a) The DEP force distribution at the oil-air interface resulting from an array of parallel electrodes, as obtained from an electrostatic calculation in accordance with Gabay et al. (2021). (b-f) The evolution of the film deformation from early times to steady state, showing the dominance of high wavenumbers at short times, and the gradual appearance of small wavenumbers with time. The analytical model successfully captures both the local and the global behaviour of the film across multiple time scales. The chamber length is 4 mm, and is thus slightly truncated on the right the by 3.8 mm field of view of the holographic microscope. In this experiment we use silicone oil with kinematic viscosity of $200\,\mathrm{cSt}$. The rest of the parameters (liquid properties and electric voltage actuation) are listed in the subsection 5.1.*

We note that subfigure 5b (representing the deformation at 0.2 ms, or $t = 5 \cdot 10^{-8}$ in non-dimensional time) does not include experimental results, as this time scale was too short for our imaging system to capture. To overcome this limitation, we conducted an additional experiment using silicone oil with a kinematic viscosity five times higher (1000 cSt) than the one presented in Figure 5 (200 cSt). Additionally, we reduced the liquid volume, resulting in a thinner film. Both modifications increased the hydrodynamic resistance of the system and consequently led to an order of magnitude slower dynamics such that t=2.44E-8 corresponds to 2ms. Despite the reduced liquid volume, the liquid remained pinned at the edges of the chamber, resulting in initial conditions where the film had a bowl shape with a height of 50 μm at the edges and approximately 20 μm at the center of the chamber. We compared the experimental measurements to an analytical solution assuming a film with a uniform height of 30 μm, which corresponds to the average film thickness in the chambers.

In subfigure 6a, we present the force distribution at the interface, characterized by 'tooth'-shaped peaks with the smallest wavelength in the system, arising from the transition region between adjacent electrodes. As discussed in section 3, large spatial wavenumbers are associated with a rapid temporal response. To isolate these features, we need to focus on short times. Subfigures 6b-f depict the deformation of the film at different time points. At 2 ms, only the highest spatial frequency is visible, in good agreement with the analytical prediction. By 22 ms, a longer wavelength becomes distinctly visible and as time progresses, it becomes the dominant feature of the deformation.

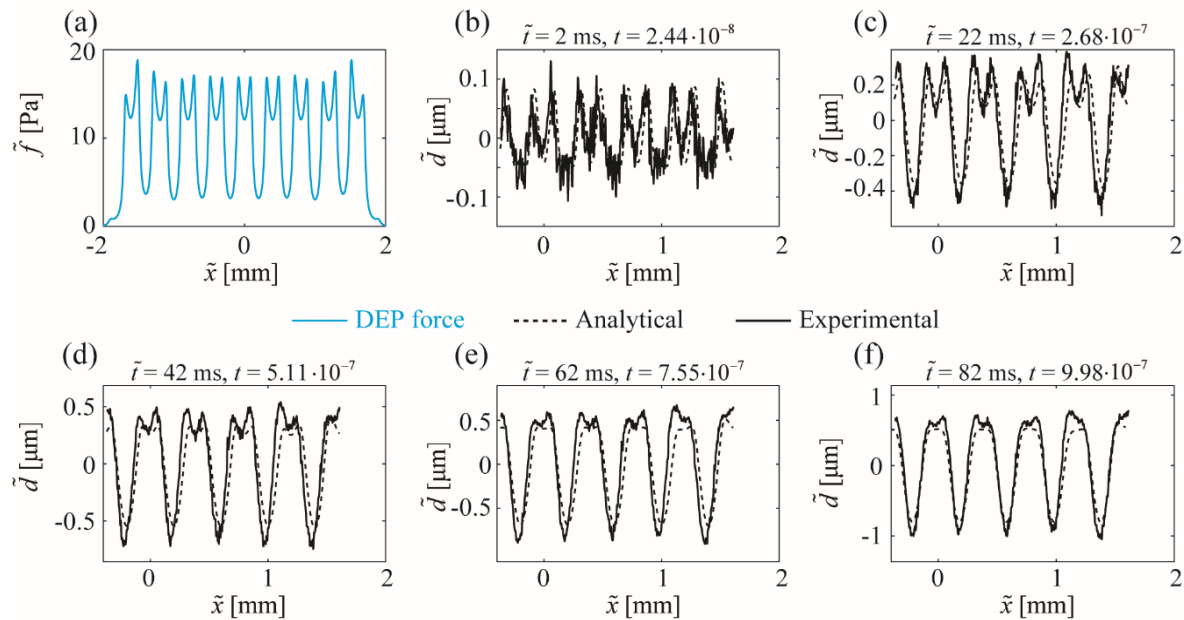

*Figure 6. Validation of the analytical model with experimental measurements at very small non-dimensional times, $t < 10^{-6}$. (a) DEP force distribution at the oil air interface of a 30 μm film, as obtained from electrostatic calculations (Gabay et al., 2021). (b-f) Comparison between the analytical solution (black dashed line) and the experimental measurements (black solid line) of the film deformations at different times. At short times the 'tooth'-shaped peaks, corresponding to the highest spatial frequency, dominate the deformation. At later times, additional lower frequencies appear and overtake the higher frequencies in amplitude. In this experiment we use silicone oil with kinematic viscosity of $1000 \, \mathrm{cSt}$. The rest of the parameters (liquid properties and electric voltage actuation) are listed in subsection 5.1.*

## 6. Concluding remarks

Thin liquid films are found in endless applications such as spin coating, self-healing, and printing technologies, where infinite boundary conditions are appropriate for describing the film dynamics. However, there is an entire class of problems that inherently rely on no-penetration boundary conditions. For example, in liquid-based actuators, such as the ones used for adaptive optics (Banerjee et al., 2018), enclosed chambers with a fixed amount of liquid are naturally the most common implementation. Banerjee *et al.* have already shown the use of a membrane suspended over a thin liquid film as a mechanism for achieving freeform optical corrections with up to $4^{th}$ degree Zernike polynomials. However, the performance of such devices could not be predicted *a priori*, as no theory was available. The model we developed here provides insight into such systems and reveals the underlying eigenfunctions and their individual and collective dynamics. For diffractive optical components, the variations in surface topography are on the order of a wavelength, and thus the solution in the small deformations regime is suitable. Clearly, for larger deformations the non-linear equation must be considered and would yield different quantitative results, but it is reasonable to expect that the boundary conditions would have a dominant effect as well. While our derivations focused on a free interface, they directly apply to the case of a membrane under dominant tension, and could also be readily extended to account for elastic bending. However, in order to quantitatively predict the response of actuation devices, the theory must be extended to two-dimensional domains.

The Closed system's boundary conditions affect both the shape of the liquid film for long time scale and the characteristic time scales required to achieve steady state. While obtaining the shape of the interface for the closed system requires solving the adjoint problem, the time scales can be readily obtained from solving the transcendental equation of the original operator. The time scale analysis shows that the closed system is 5-fold faster than the open system and 3-fold slower than the periodic one. We further showed that neither the open nor the periodic solutions can reasonably predict the deformation of the thin liquid film for general actuations. Another set of boundary conditions that is often used in the study of thin film dynamics is that of an infinite film. We did not explicitly consider this case as it's a subset of an Open system where the domain size is stretched to infinity. It thus suffers from the same issues as the open case, namely dominance of low wave number eigenfunctions ('ballooning') at long times.

In Figure 3 we presented the effect of Bond number on the eigenvalues of the of system, representing the non-dimensional time scale of the corresponding eigenfunctions. The figure shows that increasing the Bond number increases the eigenvalues and shortens the response time. However, to understand the effect of the physical parameters on the response time of the system we must consider the dimensional time scale, $\tilde{t} \propto \tilde{\tau} \lambda^{-4}$. The effect of density changes, $\tilde{\rho}$, is straightforward as it appears only in the dependence of $\lambda$ on Bo. However, an increase in surface tension $\tilde{\gamma}$ results in a decrease in time scale through $\tilde{\tau}$ and an increase in time scale through Bo. At low Bo numbers, the $\tilde{\tau}$ dependence dominates and increasing the surface tension will decrease the time scale. At large Bond numbers the increase in time scale through $\tilde{\tau}$ will balance its decrease through the Bond number and the system becomes insensitive to the value of the surface tension.

In our derivation we focused on the case of normal forces applied directly to the liquid-air interface and implemented it using DEP. However, for other actuation mechanisms such as those that dictate a slip velocity (EOF) or an interface velocity (thermocapillary flow), the differential operator and the impermeability boundary conditions would be identical. The only thing that would differ is the RHS of equation (2.4) (Rubin et al., 2017), but that would not change the solution procedure. When dictating a spatial force/velocity distribution in combination with the evolution equation, (2.2), care should be taken not to violate the underlying assumptions, at least not at the leading order of the solution; e.g. if dictating a slip velocity, then no-penetration at the domain boundaries would be fully satisfied under the lubrication approximation only if its magnitude and its spatial derivative vanish at the boundary.

Finally, we note that the non-self-adjointness is a manifestation of the mass conservation constraint on the system, and is likely to appear in other fluidic problems that are subjected to similar constraints, as well as in their numerical solutions.

## Appendix A - Analytical solutions for self-adjoint problems

We here present the solution for the classical self-adjoint problem, which we compare with our non-self-adjoint solution presented in section 2.

For simplicity, we consider the equation for small deformations for the particular case of Bo=0,

(A.1) $$d_t + d_{xxxx} = -\Pi f_{xx},$$

for an initially flat film, $d(x, t=0) = 0$.

The general solution using a Green function (Polianin, 2002) for equation (A.1) and homogenous boundary conditions is given by,

(A.2) $$d(x,t) = \int_0^2 d_0(\xi) G(x,\xi,t) d\xi + \int_0^t \int_0^2 \Phi(\xi,\tau) G(x,\xi,t-\tau) d\xi d\tau,$$

where $d_0$ is the initial deformation (zero in this case), and the source term is $\Phi(x,t) = -\Pi f_{xx}$. The Green function, $G(x,\xi,t-\tau)$ depends on the specific eigenfunctions and eigenvalues of the problem, and therefore on its boundary conditions. In the case of an 'Open chamber' the liquid is pinned at its boundaries but is free to flow in and out of the chamber. The latter condition can be expressed using the pressure relation, $p = d_{xx} - \text{Bo} \cdot d + \Pi \cdot f$, and the pinning condition, as $d_{xx}(\pm 1,t) = -\Pi f(\pm 1,t)$. Thus, the boundary conditions can be summarized by

(A.3) $$d(\pm 1,t) = 0, \quad d_{xx}(\pm 1,t) = -\Pi f(\pm 1,t).$$

For the periodic system, the boundary conditions are simply the requirement that the deformation and its derivatives up to the third order will have equal values at both edges, i.e.,

(A.4) $$d_{x,i}(-1,t) = d_{x,i}(1,t),$$

where the subscript $x$ and $i$ marks the i-th order derivative of the deformation with respect to $x$.

In order to homogenize the boundary conditions for the open case, we use a corrective polynomial function for the boundary condition, $d = v + P_o$, where

(A.5) $$P_o(x,t) = -\Pi \frac{f(1,t) + f(-1,t)}{4}(x^2 - 1) - \Pi \frac{f(1,t) - f(-1,t)}{12}(x^3 - x).$$

The Green function for the open chamber problem with homogenous boundary conditions in the domain [-1,1] is given by (Polianin, 2002)

(A.6) $$G_o(x,\xi,t) = \sum_{m=1}^{\infty} \left[ c_m \cos(\lambda_{o,m} x) \cos(\lambda_{o,m} \xi) + s_m \sin(\lambda_{o,m} x) \sin(\lambda_{o,m} \xi) \right] \exp(-\lambda_{o,m}^4 t),$$

where $c_m = \begin{cases} 1, & m = 2k-1 \\ 0, & m = 2k \end{cases}$, $s_m = \begin{cases} 0, & m = 2k-1 \\ 1, & m = 2k \end{cases}$, $k$ is a positive integer, and the eigenvalues of the system are $\lambda_{o,m} = \frac{\pi}{2} m$.

For the periodic system we obtain the Green function

(A.7) $$G_p(x,\xi,t) = \sum_{m=1}^{\infty}\left[\sin(\lambda_{p,m}x)\sin(\lambda_{p,m}\xi) + \cos(\lambda_{p,m}x)\cos(\lambda_{p,m}\xi)\right]\exp(-\lambda_{p,m}^4 t),$$

where the eigenvalues of the system are $\lambda_{p,m} = \pi m$, and the subscripts $o, p$ denote the Green functions and deformation solutions for the open and periodic systems, respectively.

Substituting (A.6) and (A.7) into (A.2), we obtain the solutions

(A.8) $$d_o(x,t) = \sum_{m=1}^{\infty}\frac{1-\exp(-\lambda_{o,m}^4 t)}{\lambda_{o,m}^4}\left[a_{o,n}\sin(\lambda_{o,m}x) + b_{o,m}\cos(\lambda_{o,m}x)\right] + P_o(x,t)$$
$$d_p(x,t) = \sum_{m=1}^{\infty}\frac{1-\exp(-\lambda_{p,m}^4 t)}{\lambda_{p,m}^4}\left[a_{p,n}\sin(\lambda_{p,m}x) + b_{p,m}\cos(\lambda_{p,m}x)\right]$$

where the series $a$ and $b$ for both systems are given by,

(A.9) $$a_{p,m} = \int_0^2 \sin(\lambda_{p,m}\xi)\Phi(\xi,\tau)d\xi, \quad b_{p,m} = \int_0^2 \cos(\lambda_{p,m}\xi)\Phi(\xi,\tau)d\xi$$
$$a_{o,m} = \int_0^2 s_n \sin(\lambda_{o,m}\xi)\Phi(\xi,\tau)d\xi, \quad b_{o,m} = \int_0^2 c_m \cos(\lambda_{o,m}\xi)\Phi(\xi,\tau)d\xi.$$

**Acknowledgments.** We gratefully acknowledge funding from the Israel Science Foundation grant no. 2263/20. I.G. acknowledges the support of ISEF and is grateful to the Azrieli Foundation for the award of an Azrieli Fellowship.

**Declaration of Interests.** The authors declare no conflict of interest.

**References:**

Backholm, M., Benzaquen, M., Salez, T., Raphaël, E. and Dalnoki-Veress, K., Capillary Levelling of a Cylindrical Hole in a Viscous Film, *Soft Matter*, vol. **10**, no. 15, pp. 2550–58, 2014. DOI: 10.1039/C3SM52940A

Banerjee, K., Rajaeipour, P., Ataman, Ç. and Zappe, H., Optofluidic Adaptive Optics, *Applied Optics*, vol. **57**, no. 22, p. 6338, August 1, 2018. DOI: 10.1364/AO.57.006338

Bénard, H., Étude Expérimentale Des Courants de Convection Dans Une Nappe Liquide.—Régime Permanent: Tourbillons Cellulaires, *Journal de Physique Théorique et Appliquée*, vol. **9**, no. 1, pp. 513–524, 1900.

Boyko, E., Eshel, R., Gommed, K., Gat, A. D. and Bercovici, M., Elastohydrodynamics of a Pre-Stretched Finite Elastic Sheet Lubricated by a Thin Viscous Film with Application to Microfluidic Soft Actuators, *Journal of Fluid Mechanics*, vol. **862**, pp. 732–52, March 2019. DOI: 10.1017/jfm.2018.967

Boyko, E., Ilssar, D., Bercovici, M. and Gat, A. D., Interfacial Instability of Thin Films in Soft Microfluidic Configurations Actuated by Electro-Osmotic Flow, *Physical Review Fluids*, vol. **5**, no. 10, p. 104201, October 1, 2020. DOI: 10.1103/PhysRevFluids.5.104201

Brown, C. V., Wells, G. G., Newton, M. I. and McHale, G., Voltage-Programmable Liquid Optical Interface, *Nature Photonics*, vol. **3**, no. 7, pp. 403–5, July 2009. DOI: 10.1038/nphoton.2009.99

Chappell, D. J. and O'Dea, R. D., Numerical-Asymptotic Models for the Manipulation of Viscous Films via Dielectrophoresis, *Journal of Fluid Mechanics*, vol. **901**, accessed September 25, 2020, from https://www.cambridge.org/core/journals/journal-of-fluid-mechanics/article/numericalasymptotic-models-for-the-manipulation-of-viscous-films-via-dielectrophoresis/31F252B5EF5E01EBA0FE660767C0654E/share/9900f9df06396b1c4554735c2d32d83bb13d21e3, October 2020. DOI: 10.1017/jfm.2020.545

Coddington, E. A. and Levinson, N., *Theory of Ordinary Differential Equations*, Tata McGraw-Hill Education, 1955.

Craster, R. V. and Matar, O. K., Dynamics and Stability of Thin Liquid Films, *Reviews of Modern Physics*, vol. **81**, no. 3, pp. 1131–98, August 5, 2009. DOI: 10.1103/RevModPhys.81.1131

Cuche, E., Marquet, P. and Depeursinge, C., Simultaneous Amplitude-Contrast and Quantitative Phase-Contrast Microscopy by Numerical Reconstruction of Fresnel off-Axis Holograms, *Applied Optics*, vol. **38**, no. 34, pp. 6994–7001, December 1, 1999. DOI: 10.1364/AO.38.006994

Dallaston, M. C., Tseluiko, D., Zheng, Z., Fontelos, M. A. and Kalliadasis, S., Self-Similar Finite-Time Singularity Formation in Degenerate Parabolic Equations Arising in Thin-Film Flows, *Nonlinearity*, vol. **30**, no. 7, pp. 2647–2666, May 2017. DOI: 10.1088/1361-6544/aa6eb3

Deissler, R. J. and Oron, A., Stable Localized Patterns in Thin Liquid Films, *Physical Review Letters*, vol. **68**, no. 19, pp. 2948–51, May 11, 1992. DOI: 10.1103/PhysRevLett.68.2948

Eshel, R., Frumkin, V., Nice, M., Luria, O., Ferdman, B., Opatovski, N., Gommed, K., Shusteff, M., Shechtman, Y. and Bercovici, M., Fabrication of Diffractive Optical Elements by Programmable Thermocapillary Shaping of Thin Liquid Films, *ArXiv:2109.00158 [Physics]*, accessed February 15, 2022, from http://arxiv.org/abs/2109.00158, August 31, 2021.

Fink, J. H. and Griffiths, R. W., Radial Spreading of Viscous-Gravity Currents with Solidifying Crust, *Journal of Fluid Mechanics*, vol. **221**, pp. 485–509, December 1990. DOI: 10.1017/S0022112090003640

Frumkin, V. and Oron, A., Liquid Film Flow along a Substrate with an Asymmetric Topography Sustained by the Thermocapillary Effect, *Physics of Fluids*, vol. **28**, no. 8, p. 082107, August 2016. DOI: 10.1063/1.4961032

Gabay, I., Paratore, F., Boyko, E., Ramos, A., Gat, A. D. and Bercovici, M., Shaping Liquid Films by Dielectrophoresis, *Flow*, vol. **1**, accessed January 30, 2022, from https://www.cambridge.org/core/journals/flow/article/shaping-liquid-films-by-dielectrophoresis/1E7D4E46CC611ED6093585E1F01283F1, ed 2021. DOI: 10.1017/flo.2021.13

Gjevik, B., Occurrence of Finite-Amplitude Surface Waves on Falling Liquid Films, *The Physics of Fluids*, vol. **13**, no. 8, pp. 1918–25, August 1970. DOI: 10.1063/1.1693186

Grotberg, J. B. and Jensen, O. E., Biofluid Mechanics in Flexible Tubes, *Annual Review of Fluid Mechanics*, vol. **36**, no. 1, pp. 121–47, 2004. DOI: 10.1146/annurev.fluid.36.050802.121918

Hewitt, I. J., Balmforth, N. J. and Bruyn, J. R. D., Elastic-Plated Gravity Currents, *European Journal of Applied Mathematics*, vol. **26**, no. 1, pp. 1–31, February 2015. DOI: 10.1017/S0956792514000291

Hosoi, A. E. and Mahadevan, L., Peeling, Healing, and Bursting in a Lubricated Elastic Sheet, *Physical Review Letters*, vol. **93**, no. 13, p. 137802, September 24, 2004. DOI: 10.1103/PhysRevLett.93.137802

Kodio, O., Griffiths, I. M. and Vella, D., Lubricated Wrinkles: Imposed Constraints Affect the Dynamics of Wrinkle Coarsening, *Physical Review Fluids*, vol. **2**, no. 1, p. 014202, January 23, 2017. DOI: 10.1103/PhysRevFluids.2.014202

Leal, L. G., *Advanced Transport Phenomena: Fluid Mechanics and Convective Transport Processes.*, Leiden: Cambridge University Press, accessed April 16, 2018, from http://public.eblib.com/choice/publicfullrecord.aspx?p=667626, 2007.

McGraw, K. O. and Wong, S. P., Forming Inferences about Some Intraclass Correlation Coefficients., *Psychological Methods*, vol. **1**, no. 1, p. 30, 1996.

Oron, A., Davis, S. H. and Bankoff, S. G., Long-Scale Evolution of Thin Liquid Films, *Reviews of Modern Physics*, vol. **69**, no. 3, pp. 931–80, July 1, 1997. DOI: 10.1103/RevModPhys.69.931

Paratore, F., Bacheva, V., Bercovici, M. and Kaigala, G. V., Reconfigurable Microfluidics, *Nature Reviews Chemistry*, vol. **6**, no. 1, pp. 70–80, 2022.

Polianin, A. D., *Handbook of Linear Partial Differential Equations for Engineers and Scientists*, Boca Raton: Chapman & Hall/CRC, pp. 781, 2002.

Rayleigh, Lord, LIX. *On Convection Currents in a Horizontal Layer of Fluid, When the Higher Temperature Is on the under Side*, The London, Edinburgh, and Dublin Philosophical Magazine and Journal of Science, vol. **32**, no. 192, pp. 529–46, December 1916. DOI: 10.1080/14786441608635602

Rubin, S., Tulchinsky, A., Gat, A. D. and Bercovici, M., Elastic Deformations Driven by Non-Uniform Lubrication Flows, *Journal of Fluid Mechanics*, vol. **812**, pp. 841–65, February 2017. DOI: 10.1017/jfm.2016.830

Simpson, J. E., Gravity Currents in the Laboratory, Atmosphere, and Ocean, *Annual Review of Fluid Mechanics*, vol. **14**, no. 1, pp. 213–34, 1982. DOI: 10.1146/annurev.fl.14.010182.001241

Williams, M. B. and Davis, S. H., Nonlinear Theory of Film Rupture, *Journal of Colloid and Interface Science*, vol. **90**, no. 1, pp. 220–28, November 1, 1982. DOI: 10.1016/0021-9797(82)90415-5

Zhao, pengpeng, Sauter, D. and Zappe, H., Tunable Fluidic Lens with Dynamic High-Order Aberration control, *Applied Optics*, accessed June 6, 2021, from https://www.osapublishing.org/ao/abstract.cfm?doi=10.1364/AO.425637, May 26, 2021. DOI: 10.1364/AO.425637

Zheng, Z., Fontelos, M. A., Shin, S., Dallaston, M. C., Tseluiko, D., Kalliadasis, S. and Stone, H. A., Healing Capillary Films, *Journal of Fluid Mechanics*, vol. **838**, pp. 404–34, March 2018. DOI: 10.1017/jfm.2017.777

Zheng, Z., Fontelos, M. A., Shin, S. and Stone, H. A., Universality in the Nonlinear Leveling of Capillary Films, *Physical Review Fluids*, vol. **3**, no. 3, p. 032001, March 26, 2018. DOI: 10.1103/PhysRevFluids.3.032001